\documentclass[a4paper,UKenglish,cleveref, autoref, thm-restate]{lipics-v2021}
\pdfoutput=1 %uncomment to ensure pdflatex processing (mandatatory e.g. to submit to arXiv)
\hideLIPIcs  %uncomment to remove references to LIPIcs series (logo, DOI, ...), e.g. when preparing a pre-final version to be uploaded to arXiv or another public repository

\title{Can LLMs Build a MaxSAT Solver from Papers? The CoreForge Experience} %TODO Please add

\author{Ruben Martins}{Carnegie Mellon University \and \url{https://www.cs.cmu.edu/~rubenm/} }{rubenm@andrew.cmu.edu}{https://orcid.org/0000-0003-1525-1382}{}

\authorrunning{Ruben Martins} 

\Copyright{Ruben Martins} 

\ccsdesc[500]{Software and its engineering~Automatic programming}
\ccsdesc[500]{Theory of computation~Automated reasoning}
\ccsdesc[300]{Theory of computation~Constraint and logic programming}

\keywords{MaxSAT, LLMs, Code Generation} 

\relatedversion{} 

\funding{This work was partially supported by the National Science Foundation (NSF) under Award CCF2427581 and DARPA under Agreement FA8750-24-9-1000.}

\nolinenumbers %uncomment to disable line numbering

\EventEditors{John Q. Open and Joan R. Access}
\EventNoEds{2}
\EventLongTitle{42nd Conference on Very Important Topics (CVIT 2016)}
\EventShortTitle{CVIT 2016}
\EventAcronym{CVIT}
\EventYear{2016}
\EventDate{December 24--27, 2016}
\EventLocation{Little Whinging, United Kingdom}
\EventLogo{}
\SeriesVolume{42}
\ArticleNo{23}
\usepackage{xspace}
\usepackage{tikz}
\usepackage{amssymb}
\usepackage{booktabs}
\usetikzlibrary{arrows.meta, positioning, fit, backgrounds}
\newcommand{\coreforge}{{\textsf CoreForge}\xspace}

\begin{document}

\maketitle

\begin{abstract}
We report on \coreforge, an experience in using large language models (LLMs)
to build an unweighted MaxSAT solver from research papers rather than from an
existing solver codebase. The project focuses on unsatisfiability-based MaxSAT
algorithms and follows an iterative workflow that combines paper discussions
with ChatGPT, implementation through Codex prompts, and repeated LLM-assisted
code audits and revisions. Although the codebase implements several algorithms
and solver components, our evaluation focuses on configurations that combine
core-guided optimization, lightweight preprocessing, core minimization,
integration with integer linear optimization backends, and a new core-sequence lookahead
approach.

Our experience suggests that LLMs can support solver implementation
from papers, while requiring external validation, benchmarking, and human
guidance. In our experiments, fuzzing and MaxSAT Evaluation instances did not
reveal wrong answers in the tested configurations, although performance remains
below the best hand-engineered MaxSAT solvers. We summarize what worked, what remained difficult, and the lessons for future
LLM-assisted solver development.
\end{abstract}

\section{Introduction}

Large language models (LLMs) can generate substantial software artifacts, but
their ability to support the development of constraint solvers
remains less clear. Solver development is a demanding test
case because competitive solvers combine algorithmic ideas with careful
implementation, preprocessing, heuristics, data structures, and extensive
performance tuning.

Recent work has explored LLMs for solver engineering, including heuristic
optimization for existing SAT solvers, repository-scale SAT solver evolution,
SAT-solving code generation, and SMT solver construction~\cite{autosat2024,
satlution2025,solsearch2025,automodsat2025,llm2smt2026}. In parallel,
paper-to-code benchmarks and systems study whether AI agents can implement
research ideas from papers~\cite{paperbench2025,paper2code2025,
researchcodebench2025}. \coreforge connects these directions by studying whether 
LLMs can build a MaxSAT solver from research papers rather than from an existing solver
codebase.

This distinction is important because most solver-focused systems either start from existing SAT solver implementations, optimize selected components, or target restricted SMT fragments. In contrast, \coreforge studies a longer-horizon paper-to-code task in which ideas from multiple MaxSAT papers are turned into a growing solver codebase, revised over many iterations, and evaluated on standard MaxSAT benchmarks. The result is not a solver-evolution study over an existing implementation, but a case study in iterative LLM-assisted solver construction from papers.

This experience addresses three questions: (1) can LLMs build a nontrivial MaxSAT
solver from papers, (2) were wrong answers observed,
and (3) how far is the resulting performance from hand-engineered MaxSAT solvers?

\section{The \coreforge Experience}

\coreforge was developed as an iterative paper-to-code experiment, not a
one-shot generation task. Each iteration translated MaxSAT ideas into prompts,
integrated generated changes, and used audits, fuzzing, and benchmarks to guide
revisions.

\vspace{-4mm}
\paragraph*{Can LLMs build a MaxSAT solver from papers?}
\vspace{-2mm}
In our experience, yes, but not in a one-shot or fully autonomous way.
\coreforge was built from scratch rather than from an existing MaxSAT solver
codebase. The human role was to select papers, define implementation goals,
run experiments, and decide which generated changes to keep. The LLM produced
the main solver architecture and implemented several unsatisfiability-based
MaxSAT algorithms, including PM2~\cite{pm2}, MSU3~\cite{msu3}, and
OLL~\cite{oll}, together with lightweight preprocessing~\cite{rc2}, core
minimization~\cite{rc2}, integration with optimization backends such as SCIP~\cite{scip}
and CP-SAT~\cite{cp-sat}, and a new core-sequence lookahead approach inspired by recent work on OLL
reformulation structure and core sequences~\cite{schidler2025dag,
alos2026coresequences}. The solver runs bounded probes with different
core-extraction strategies, scores the resulting core-sequence prefixes, and then
commits to the most promising strategy for the main search. The final codebase
contains more than 25K lines of C++ code and includes algorithms and components
beyond those enabled in our evaluation. However, the process required repeated
prompting, benchmarking, and LLM-assisted auditing. The LLM was effective at
translating high-level algorithmic ideas into code, but less effective at
aggressive low-level solver engineering.
Appendix~\ref{app:workflow} gives additional details on the development
workflow.

\vspace{-4mm}
\paragraph*{Were wrong answers observed?}
\vspace{-2mm}
We evaluated the tested configurations using fuzzing~\cite{paxian2023fuzzing} and MaxSAT Evaluation instances, and did not observe wrong answers. This was surprising given the size of the generated codebase and the fact that we intentionally treated the implementation largely as a black box, relying on testing, benchmarking, and LLM-assisted audits rather than detailed manual code inspection. This evidence should be interpreted cautiously. During development, fuzzing and
LLM-assisted audits uncovered issues that were later fixed by the coding agent, underscoring the
need for external validation.

\vspace{-4mm}
\paragraph*{How competitive are the solvers?}
\vspace{-2mm}

\coreforge{} solves a substantial number of nontrivial MaxSAT Evaluation
instances, although it is not yet competitive with the best hand-engineered
MaxSAT solvers. We evaluated three configurations: a core-guided OLL baseline,
a variant with integer-linear-optimization backend integration, and a variant
with core-sequence lookahead. While the individual configurations remain below
the strongest solvers, they outperform several established solvers on our
benchmark subset. Their virtual best solver suggests that better instance-level
selection could further improve performance. Appendices~\ref{app:configs}
and~\ref{app:performance} give more details.

\section{Conclusion and Future Work}

\coreforge suggests that LLMs can support substantial solver construction from
research papers when paired with iterative guidance, benchmarking, and external
validation. The main lesson is that LLM-assisted paper-to-code workflows can already support
solver development, but trust and autonomy remain central challenges.

Future work should study more autonomous, self-improving solver agents that
can read papers, propose implementations, run benchmarks, audit failures, and
revise their own code. A central challenge is ensuring that they implement the
intended algorithms, rather than producing code that merely works on the
available benchmarks.

\bibliography{bibtex}

@article{autosat2024,
  author       = {Yiwen Sun and
                  Furong Ye and
                  Xianyin Zhang and
                  Shiyu Huang and
                  Bingzhen Zhang and
                  Ke Wei and
                  Shaowei Cai},
  title        = {{AutoSAT: Automatically Optimize {SAT} Solvers via Large Language Models}},
  journal      = {CoRR},
  volume       = {abs/2402.10705},
  year         = {2024},
  eprinttype   = {arXiv},
  eprint       = {2402.10705},
}

@article{satlution2025,
  author       = {Cunxi Yu and
                  Rongjian Liang and
                  Chia{-}Tung Ho and
                  Haoxing Ren},
  title        = {{Autonomous Code Evolution Meets NP-Completeness}},
  journal      = {CoRR},
  volume       = {abs/2509.07367},
  year         = {2025},
  eprinttype   = {arXiv},
  eprint       = {2509.07367},
}

@inproceedings{solsearch2025,
  author       = {Junjie Sheng and
                  Yanqiu Lin and
                  Jiehao Wu and
                  Yanhong Huang and
                  Jianqi Shi and
                  Min Zhang and
                  Xiangfeng Wang},
  title        = {{SolSearch: An LLM-Driven Framework for Efficient SAT-Solving Code
                  Generation}},
  booktitle    = {International Conference on Software Engineering:
                  New Ideas and Emerging Results, {ICSE} 2025 - NIER},
  pages        = {6--10},
  publisher    = {{IEEE}},
  year         = {2025},
}

@article{automodsat2025,
  author       = {Yiwen Sun and
                  Furong Ye and
                  Zhihan Chen and
                  Ke Wei and
                  Shaowei Cai},
  title        = {Automatically discovering heuristics in a complex {SAT} solver with
                  large language models},
  journal      = {CoRR},
  volume       = {abs/2507.22876},
  year         = {2025},
  eprinttype   = {arXiv},
  eprint       = {2507.22876},
}

@article{paper2code2025,
  author       = {Minju Seo and
                  Jinheon Baek and
                  Seongyun Lee and
                  Sung Ju Hwang},
  title        = {{Paper2Code: Automating Code Generation from Scientific Papers in Machine
                  Learning}},
  journal      = {CoRR},
  volume       = {abs/2504.17192},
  year         = {2025},
  eprinttype   = {arXiv},
  eprint       = {2504.17192},
}

@article{researchcodebench2025,
  author       = {Tianyu Hua and
                  Harper Hua and
                  Violet Xiang and
                  Benjamin Klieger and
                  Sang T. Truong and
                  Weixin Liang and
                  Fan{-}Yun Sun and
                  Nick Haber},
  title        = {{ResearchCodeBench: Benchmarking LLMs on Implementing Novel Machine
                  Learning Research Code}},
  journal      = {CoRR},
  volume       = {abs/2506.02314},
  year         = {2025},
  eprinttype   = {arXiv},
  eprint       = {2506.02314},
}

@article{llm2smt2026,
  author       = {Mikol{\'{a}}s Janota and
                  Mirek Ols{\'{a}}k},
  title        = {{{LLM2SMT:} Building an {SMT} Solver with Zero Human-Written Code}},
  journal      = {CoRR},
  volume       = {abs/2603.06931},
  year         = {2026},
  eprinttype   = {arXiv},
  eprint       = {2603.06931},
}

@article{paperbench2025,
  author       = {Giulio Starace and
                  Oliver Jaffe and
                  Dane Sherburn and
                  James Aung and
                  Jun Shern Chan and
                  Leon Maksin and
                  Rachel Dias and
                  Evan Mays and
                  Benjamin Kinsella and
                  Wyatt Thompson and
                  Johannes Heidecke and
                  Amelia Glaese and
                  Tejal Patwardhan},
  title        = {{PaperBench: Evaluating AI's Ability to Replicate {AI} Research}},
  journal      = {CoRR},
  volume       = {abs/2504.01848},
  year         = {2025},
  eprinttype   = {arXiv},
  eprint       = {2504.01848},
}

@inproceedings{pm2,
  author    = {Carlos Ans{\'o}tegui and Maria Luisa Bonet and Jordi Levy},
  title     = {Solving Weighted Partial MaxSAT through Satisfiability Testing},
  booktitle = {Proceedings of the 12th International Conference on Theory and Applications of Satisfiability Testing (SAT)},
  series    = {Lecture Notes in Computer Science},
  volume    = {5584},
  pages     = {427--440},
  publisher = {Springer},
  year      = {2009}
}

@article{rc2,
  author  = {Alexey Ignatiev and Ant{\'o}nio Morgado and Joao Marques-Silva},
  title   = {{RC2}: An Efficient MaxSAT Solver},
  journal = {Journal on Satisfiability, Boolean Modeling and Computation},
  volume  = {11},
  number  = {1},
  pages   = {53--64},
  year    = {2019},
  doi     = {10.3233/SAT190116}
}

@inproceedings{msu3,
  author    = {Ruben Martins and Saurabh Joshi and Vasco Manquinho and In{\^e}s Lynce},
  title     = {Incremental Cardinality Constraints for MaxSAT},
  booktitle = {Proceedings of the 20th International Conference on Principles
               and Practice of Constraint Programming (CP)},
  series    = {Lecture Notes in Computer Science},
  volume    = {8656},
  pages     = {531--548},
  publisher = {Springer},
  year      = {2014},
  doi       = {10.1007/978-3-319-10428-7_39}
}

@inproceedings{oll,
  author    = {Ant{\'o}nio Morgado and Carmine Dodaro and Joao Marques-Silva},
  title     = {Core-Guided MaxSAT with Soft Cardinality Constraints},
  booktitle = {Proceedings of the 20th International Conference on Principles
               and Practice of Constraint Programming (CP)},
  series    = {Lecture Notes in Computer Science},
  volume    = {8656},
  pages     = {564--573},
  publisher = {Springer},
  year      = {2014},
  doi       = {10.1007/978-3-319-10428-7_42}
}

@article{alos2026coresequences,
  author  = {Josep Al{\`o}s and Carlos Ans{\'o}tegui and Eduard Torres},
  title   = {Revisiting SAT-Based Solvers: MaxSAT Rules and Core Sequences},
  journal = {Journal of Artificial Intelligence Research},
  volume  = {85},
  pages   = {Article 16},
  year    = {2026},
  doi     = {10.1613/jair.1.19525}
}

@inproceedings{schidler2025dag,
  author    = {Andr{\'e} Schidler and Stefan Szeider},
  title     = {Analyzing Reformulation Performance in Core-Guided MaxSAT Solving},
  booktitle = {28th International Conference on Theory and Applications of Satisfiability Testing (SAT 2025)},
  series    = {Leibniz International Proceedings in Informatics (LIPIcs)},
  volume    = {341},
  pages     = {26:1--26:18},
  publisher = {Schloss Dagstuhl -- Leibniz-Zentrum f{\"u}r Informatik},
  year      = {2025},
  doi       = {10.4230/LIPIcs.SAT.2025.26}
}

@article{scip,
  author  = {Tobias Achterberg},
  title   = {{SCIP}: Solving Constraint Integer Programs},
  journal = {Mathematical Programming Computation},
  volume  = {1},
  number  = {1},
  pages   = {1--41},
  year    = {2009},
  doi     = {10.1007/s12532-008-0001-1}
}

@inproceedings{cp-sat,
  author    = {Laurent Perron and Vincent Furnon},
  title     = {The {CP-SAT-LP} Solver},
  booktitle = {29th International Conference on Principles and Practice of Constraint Programming (CP 2023)},
  series    = {Leibniz International Proceedings in Informatics (LIPIcs)},
  volume    = {280},
  pages     = {3:1--3:2},
  publisher = {Schloss Dagstuhl -- Leibniz-Zentrum f{\"u}r Informatik},
  year      = {2023},
  doi       = {10.4230/LIPIcs.CP.2023.3}
}

@inproceedings{paxian2023fuzzing,
  author    = {Tobias Paxian and Armin Biere},
  title     = {Uncovering and Classifying Bugs in MaxSAT Solvers through Fuzzing and Delta Debugging},
  booktitle = {Proceedings of the 14th International Workshop on Pragmatics of SAT (POS 2023)},
  series    = {CEUR Workshop Proceedings},
  volume    = {3545},
  pages     = {59--71},
  publisher = {CEUR-WS.org},
  year      = {2023}
}

@inproceedings{neves2015resolution,
  author    = {Miguel Neves and Ruben Martins and Mikol{\'a}{\v{s}} Janota and In{\^e}s Lynce and Vasco Manquinho},
  title     = {Exploiting Resolution-Based Representations for MaxSAT Solving},
  booktitle = {Proceedings of the 18th International Conference on Theory and Applications of Satisfiability Testing (SAT)},
  series    = {Lecture Notes in Computer Science},
  volume    = {9340},
  pages     = {272--286},
  publisher = {Springer},
  year      = {2015},
  doi       = {10.1007/978-3-319-24318-4_20}
}

\appendix

\section{Development Workflow}
\label{app:workflow}

\begin{figure}[t]
\centering
\begin{tikzpicture}[
    font=\small,
    >=Latex,
    box/.style={
        draw=black!55,
        rounded corners=4pt,
        align=center,
        minimum width=2.6cm,
        minimum height=0.85cm,
        fill=black!3
    },
    arrow/.style={->, thick, draw=black!70}
]

\node[box] (papers) at (0,0) {Papers};
\node[box] (planning) at (3.2,0) {LLM Planning};
\node[box] (coding) at (6.4,0) {LLM Coding};
\node[box] (audit) at (6.4,-1.8) {Audit \& Revise};
\node[box] (eval) at (3.2,-1.8) {Evaluate};

\draw[arrow] (papers) -- (planning);
\draw[arrow] (planning) -- (coding);
\draw[arrow] (coding) -- (audit);
\draw[arrow] (audit) -- (eval);
\draw[arrow] (eval) -- node[left, font=\scriptsize] {iterate} (planning);

\end{tikzpicture}
\caption{High-level iterative workflow used to develop \coreforge. Papers were
used for planning, planning was converted into implementation prompts, and the
resulting code was evaluated, audited, and revised across iterations.}
\label{fig:workflow}
\end{figure}
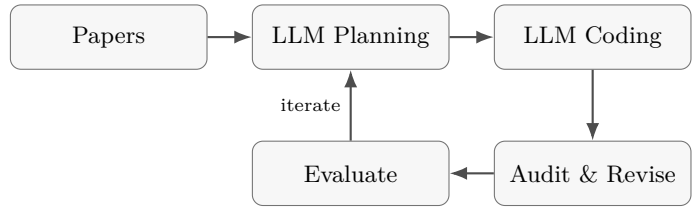

This appendix summarizes the workflow used to develop \coreforge. The goal was
to study whether a MaxSAT solver could be built from research papers, rather
than from an existing solver codebase. Accordingly, research papers were the
primary technical input, and source code from other MaxSAT solvers was not
analyzed.

Figure~\ref{fig:workflow} shows the overall workflow. The process began by
providing papers to ChatGPT one by one, using the discussions to extract the
main algorithmic ideas, identify promising features, and plan implementation
steps. These planning discussions were then converted into concrete prompts for
Codex, which generated or revised solver code. The resulting code was audited
with ChatGPT and Codex to identify mismatches between the intended algorithm
and the implementation, missing corner cases, and opportunities for revision.
This process was repeated throughout development.

A practical motivation for separating planning from coding was cost control.
ChatGPT was used primarily for higher-level planning and design discussions,
while Codex was reserved for implementation tasks. Fuzzing and benchmark
evaluation were run manually, rather than delegated to the agent, to keep
control over the validation process and avoid spending Codex time on long
experimental runs.

The development spanned approximately four weeks and largely exhausted the
weekly Codex usage budget in the \$20 plan. Most of the work used GPT-5.4,
while GPT-5.5 became available near the end of the project. Overall, the
process involved more than 70 large prompts, around 120 Git commits, and
analysis of roughly 30 MaxSAT papers together with MaxSAT Evaluation
proceedings. The commits were created manually after incorporating
LLM-generated changes. No external solver code was analyzed, since the purpose
of the experiment was to study a paper-to-code workflow.

The difficulty of implementation varied substantially across features. Some
components, such as incremental totalizer encodings~\cite{msu3}, required many iterations
before they were integrated correctly. Other ideas were implemented only
partially or differently from the original papers, as happened with some
partitioning-based techniques~\cite{neves2015resolution}. As the codebase grew, integration also became harder. Adding new features,
refactoring existing components, and removing unused code increasingly consumed
a significant fraction of the agent's effort.

In addition to code generation, the LLMs were used for implementation audits.
This was important because we did not manually inspect the generated code.
Instead, we treated the solver largely as a black box and relied on
LLM-assisted audits, manually run fuzzing, and manually run benchmark
evaluation to identify correctness and performance issues.

\section{New Feature Development: Core-Sequence Lookahead}
\label{app:lookahead}

One part of the \coreforge development process went beyond reimplementing
existing algorithms and explored whether LLMs could help implement new
features. The resulting approach, called \emph{core-sequence
lookahead}, was motivated by recent work showing that the trajectory of a
core-guided MaxSAT solver matters. Schidler and Szeider analyze OLL
reformulations through the structure of the generated cores~\cite{schidler2025dag},
while Al{\`o}s, Ans{\'o}tegui, and Torres show that different core sequences
can lead to substantially different solving behavior~\cite{alos2026coresequences}.
Inspired by these observations, core-sequence lookahead explores several early
core sequences before committing to the main search.

The goal of core-sequence lookahead is to use a small, bounded probing phase to
compare alternative early search trajectories before committing to the main
run. Each probe follows a different strategy for ordering assumptions and
extracting or processing cores. The solver records the resulting core-sequence
prefix, scores it using inexpensive statistics, and selects the most promising
strategy for the subsequent search. Thus, this approach turns variation in early
core discovery into an explicit, bounded selection step.

\begin{figure}[t]
\centering
\begin{tikzpicture}[
    font=\small,
    >=Latex,
    box/.style={
        draw=black!55,
        rounded corners=4pt,
        align=center,
        minimum width=2.65cm,
        minimum height=0.82cm,
        fill=black!3
    },
    arrow/.style={->, thick, draw=black!70}
]

\node[box] (formula) at (0,0) {Input formula};

\node[box] (p1) at (3.25,1.45) {Probe 1};
\node[box] (p2) at (3.25,0.48) {Probe 2};
\node       (dots) at (3.25,-0.22) {$\vdots$};
\node[box] (pk) at (3.25,-1.20) {Probe $k$};

\node[box] (score) at (6.75,0.10) {Score prefixes};
\node[box] (commit) at (10.05,0.10) {Commit to\\selected strategy};

\draw[arrow] (formula.east) -- (p1.west);
\draw[arrow] (formula.east) -- (p2.west);
\draw[arrow] (formula.east) -- (pk.west);

\draw[arrow] (p1.east) -- (score.west);
\draw[arrow] (p2.east) -- (score.west);
\draw[arrow] (pk.east) -- (score.west);

\draw[arrow] (score.east) -- (commit.west);

\end{tikzpicture}
\caption{High-level structure of core-sequence lookahead. Several bounded
probes explore alternative early core sequences; the solver scores the
observed prefixes and commits to the most promising strategy.}
\label{fig:lookahead}
\end{figure}
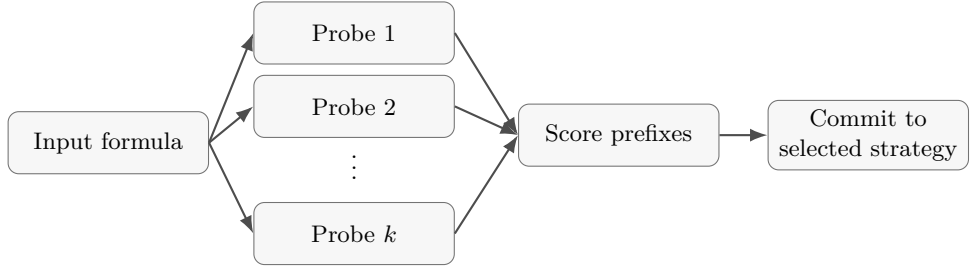

Figure~\ref{fig:lookahead} shows the high-level structure. The solver first
runs several bounded probes. Each probe attempts to extract a short sequence of
cores under a SAT-call and conflict budget. The resulting sequences are scored
using inexpensive statistics, such as lower-bound gain, core sizes, and the
number of distinct soft clauses involved. The selected strategy is then used for
the subsequent MaxSAT search.

The feature was implemented through the same LLM-assisted workflow used for the
rest of the solver. The high-level idea was discussed with ChatGPT and then
converted into a Codex implementation prompt. Codex added the lookahead
infrastructure, including strategy definitions, probe budgets, prefix statistics,
scoring, and integration with the main solver configuration. Subsequent
LLM-assisted audits focused on implementation risks specific to lookahead,
including whether probe state was kept separate from the committed run,
whether budgets were enforced consistently, and whether the selected strategy was
actually used after probing.

In the evaluated lookahead configuration, \coreforge runs eight bounded probes
before the main MaxSAT search. Each probe has limits on the number of SAT
calls, conflicts per call, and cores collected. After this bounded exploration, the solver selects the most promising strategy and restarts the main search using that strategy.

This appendix does not present core-sequence lookahead as a finished MaxSAT
technique. As discussed in the evaluation (Appendix~\ref{app:performance}), the current implementation still has
limitations. Instead, the feature serves as an example of a different use of
LLMs in solver development. The LLM was not only asked to implement known
algorithms from papers, but also to help turn a research intuition into a
concrete solver feature, integrate it into a growing codebase, and revise it
through audit and benchmarking.

\begin{table}[t]
\centering
\caption{\coreforge components in the three evaluation configurations.}
\small
\begin{tabular}{lccc}
\toprule
\textbf{Component} & \textbf{Baseline} & \textbf{ILP} & \textbf{Lookahead} \\
\midrule
OLL Core-guided Algorithm~\cite{oll}              & \checkmark & \checkmark & \checkmark \\
Intrinsic AtMost1 preprocessing~\cite{rc2} & \checkmark & \checkmark & \checkmark \\
Disjoint-core preprocessing~\cite{pm2}     & \checkmark & \checkmark & \checkmark \\
Core minimization~\cite{rc2}               & \checkmark & \checkmark & \checkmark \\
Initial upper bound             &            & \checkmark & \checkmark \\
SCIP presolve/search~\cite{scip}            &            & \checkmark & \checkmark \\
CP-SAT backend~\cite{cp-sat}                  &            & \checkmark & \checkmark \\
Core-sequence lookahead         &            &            & \checkmark \\
\bottomrule
\end{tabular}
\label{tab:eval-configs}
\end{table}

\section{Evaluation Configurations}
\label{app:configs}

This appendix summarizes the three \coreforge configurations used in the
evaluation. The configurations are summarized in Table~\ref{tab:eval-configs}.
All configurations target unweighted MaxSAT and share the same core-guided
MaxSAT infrastructure. These configurations are being prepared for submission to the MaxSAT Evaluation 2026.

\begin{itemize}
\item \textbf{Baseline.}
The baseline configuration measures the performance of the LLM-generated
MaxSAT solver without external optimization backends. It uses an OLL-style
unsatisfiabi\-li\-ty-based algorithm~\cite{oll}, with incremental construction of
cardinality constraints~\cite{msu3}, as the main optimization procedure. The
configuration also enables lightweight preprocessing, including intrinsic
AtMost1 reasoning over soft clauses~\cite{rc2}, disjoint-core
preprocessing~\cite{pm2}, and bounded core minimization~\cite{rc2}. 
%This
%configuration serves as the reference point for evaluating the generated
%MaxSAT implementation without the overhead of additional backend or lookahead
%components.

\item \textbf{ILP.}
The ILP configuration extends the baseline with optimization-backend
integration. It first computes an initial upper bound by finding a feasible
model and then applying a bounded local-improvement phase that perturbs the
model and keeps assignments that reduce the number of unsatisfied soft clauses.
The resulting upper bound is passed to the subsequent optimization stages. The
configuration also runs SCIP~\cite{scip} for presolving and solving, and
CP-SAT~\cite{cp-sat} as an additional optimization backend, each with a short
timeout. When these backend stages do not solve the instance, the remaining
search is handled by the core-guided MaxSAT engine.

\item \textbf{Lookahead.}
The lookahead configuration extends the ILP configuration with core-sequence
lookahead, described in Appendix~\ref{app:lookahead}. Before the main MaxSAT
search, the solver runs a bounded probing phase with several strategies for
ordering assumptions and processing early cores. The observed prefixes are
scored using lightweight statistics, and the selected strategy is used for the
subsequent search.
\end{itemize}

\section{Performance Results}
\label{app:performance}

\newcommand{\mse}{\textsuperscript{\textsc{MSE'24}}}
\newcommand{\cf}{\textsuperscript{\textsc{CF}}}

This appendix reports the performance of the three \coreforge{} configurations
used in our evaluation. The goal of these experiments is not to claim that
\coreforge{} improves the state of the art, but to assess the current
implementation, identify its main bottlenecks, and quantify the potential impact
of the proposed extensions.

\vspace{-4mm}
\paragraph*{Experimental setup}
\vspace{-2mm}
We evaluated the baseline, ILP, and lookahead configurations on a
417-instance subset of the MaxSAT Evaluation 2024 benchmarks. Since the goal is
a diagnostic comparison rather than a full competition-scale evaluation, we focus
on instances that were solved by at least one solver in the MaxSAT Evaluation
2024 and whose optimum value is larger than zero. This selection gives a
nontrivial benchmark subset with known reference optima, while avoiding
instances that are either unsolved by all evaluated solvers or have a zero-cost
optimum.

\begin{figure}[t]
\centering
% Replace with the generated plot.
\includegraphics[width=0.9\linewidth]{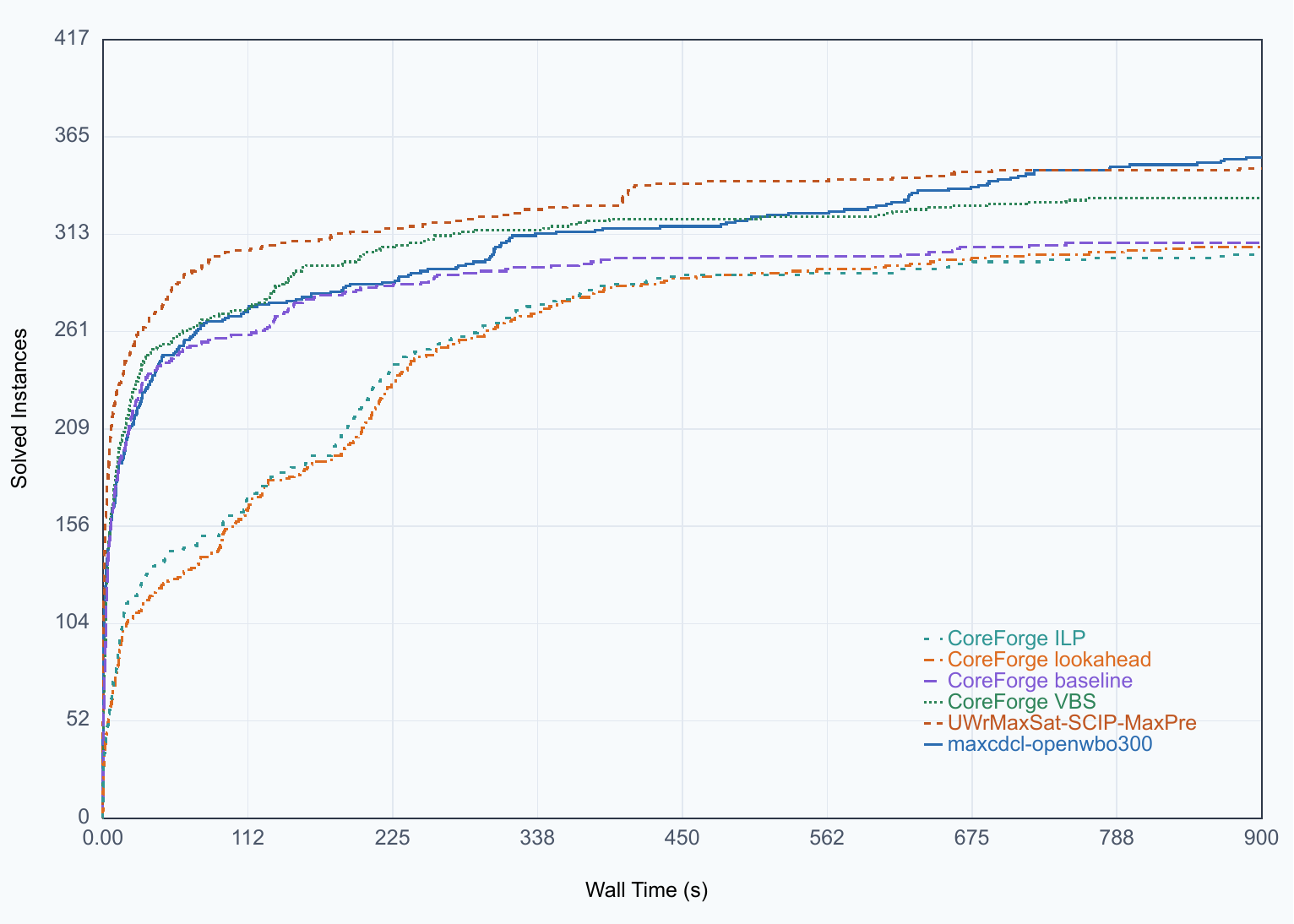}
\caption{Cactus plot comparing the baseline, ILP, and lookahead configurations, and their VBS against two strong solvers from the unweighted track of the MaxSAT Evaluation 2024.}
\label{fig:performance-cactus}
\end{figure}

\paragraph*{Overall performance}
Figure~\ref{fig:performance-cactus} shows a cactus plot comparing the three
\coreforge{} configurations and their VBS against \texttt{maxcdcl-openwbo300}
and \texttt{UWrMaxSat-SCIP-MaxPre}. The plot reports, for each solver, the
number of instances solved within a given runtime budget.

The cactus plot shows that the current \coreforge{} configurations still lag
behind the strongest solvers, both in the number of solved instances and in the
rate at which instances are solved. This is expected, since \coreforge{} is still
an early implementation and several of its components are not yet guarded by
robust instance-level resource policies. In particular, the ILP and lookahead
configurations can introduce overhead on instances where their additional
reasoning is not beneficial. This explains why these configurations do not
strictly dominate the baseline, even though they solve additional instances on
some benchmark families.

Beyond the aggregate runtime curves, the per-instance comparison shows that
\coreforge{} is not merely solving a subset of the instances solved by existing
solvers. The \coreforge{} VBS solves instances missed by strong baselines such
as \texttt{UWrMaxSat-SCIP-MaxPre} and \texttt{maxcdcl}. In particular,
\coreforge{} solves 9 instances that are not solved by
\texttt{UWrMaxSat-SCIP-MaxPre} and 19 instances that are not solved by
\texttt{maxcdcl}.

Conversely, the instances solved by these solvers but missed by \coreforge{}
highlight the main remaining weaknesses of the current implementation.
\texttt{UWrMaxSat-SCIP-MaxPre} solves 25 instances that are not solved by
\coreforge{}, 13 of which belong to the \texttt{frb} family. Similarly,
\texttt{maxcdcl} solves 41 instances that are not solved by \coreforge{}, with
35 of them concentrated in two families, namely 13 in \texttt{frb} and 22 in
\texttt{minimize-5gons}. This concentration suggests that the gap to the
strongest solvers is not uniformly spread across the benchmark set. Instead, a
small number of benchmark families account for a large fraction of the remaining
gap. Improving \coreforge{} on these families could therefore bring its
performance substantially closer to the state-of-the-art solvers.

\begin{table}[t]
\centering
\caption{Number of instances solved within the timeout on the selected
417-instance subset.}
\label{tab:solver-results}
\small
\setlength{\tabcolsep}{5pt}
\begin{tabular}{@{}lcr@{}}
\toprule
\textbf{Solver} & \textbf{Source} & \textbf{Solved} \\
\midrule
\texttt{maxcdcl-openwbo300}        & CF     & 354 \\
\texttt{maxcdcl-openwbo300}        & MSE'24 & 354 \\
\texttt{maxcdcl}                   & CF     & 348 \\
\texttt{UWrMaxSat-SCIP-MaxPre}     & CF     & 348 \\
\texttt{UWrMaxSat-SCIP-MaxPre}     & MSE'24 & 341 \\
\texttt{EvalMaxSAT-SBVA}           & MSE'24 & 340 \\
\coreforge{} VBS                   & CF     & 332 \\
\texttt{EvalMaxSAT-SBVA-saveCores} & MSE'24 & 329 \\
\texttt{EvalMaxSAT-SBVA\_1}        & MSE'24 & 328 \\
\texttt{wmaxcdcl}                  & MSE'24 & 317 \\
\texttt{wmaxcdcl-openwbo1200}      & MSE'24 & 316 \\
\coreforge{} baseline              & CF     & 308 \\
\coreforge{} lookahead             & CF     & 306 \\
\coreforge{} ILP                   & CF     & 302 \\
\texttt{Pacose}                    & MSE'24 & 292 \\
\texttt{PacoseMP2}                 & MSE'24 & 288 \\
\texttt{Exact}                     & MSE'24 & 274 \\
\texttt{CASHWMaxSAT-DisjCad-S15}   & MSE'24 & 248 \\
\texttt{CASHWMaxSAT-DisjCad-S18}   & MSE'24 & 248 \\
\texttt{CASHWMaxSAT-DisjCom-S15}   & MSE'24 & 240 \\
\texttt{CASHWMaxSAT-DisjCom-S18}   & MSE'24 & 240 \\
\bottomrule
\end{tabular}

\vspace{0.5em}
\footnotesize
CF results were obtained on our local machines, each with 32 cores, 64\,GB RAM,
and an Intel Xeon Silver 4110 CPU at 2.10\,GHz. Local runs used a 16\,GB memory
limit and a 900\,s timeout. MSE'24 results are taken from the MaxSAT Evaluation
2024, which used a 1200\,s timeout. Since the results were obtained on different
machines and with different time limits, the comparison should be interpreted as
approximate rather than as a fully controlled head-to-head evaluation. The
\coreforge{} VBS row is the virtual best over the baseline, ILP, and lookahead
configurations.
\end{table}

\vspace{-4mm}
\paragraph*{Comparison against other MaxSAT Evaluation 2024 solvers}
\vspace{-2mm}
Table~\ref{tab:solver-results} compares \coreforge{} against the strongest
solvers on the selected 417-instance subset. The results should be interpreted
with care because they combine runs obtained in two environments. Some solvers
were rerun on our local machines, while the remaining results are taken from the
MaxSAT Evaluation 2024. The locally rerun baselines provide a useful calibration
point. In particular, \texttt{maxcdcl-openwbo300} solves the same number of
instances locally as in the evaluation data, even though our runs used a
900\,s timeout while the MaxSAT Evaluation 2024 used a 1200\,s timeout.
Moreover, \texttt{UWrMaxSat-SCIP-MaxPre} solves slightly more instances in our
local setting. These results suggest that the local runs are useful for
identifying broad performance trends, although the comparison is not a fully
controlled head-to-head evaluation.

The current individual \coreforge{} configurations are not yet competitive with
the best solvers in total number of solved instances. The baseline solves 308
instances, while the lookahead and ILP configurations solve 306 and 302
instances, respectively. However, the \coreforge{} virtual best solver (VBS)
solves 332 instances, substantially more than any individual \coreforge{}
configuration. This gap shows that the three configurations are complementary,
since each configuration solves instances that the others miss. It also suggests
that much of the gap to the VBS is recoverable.

The main limitation is overhead control. The ILP and lookahead configurations
extend the baseline, but in the current implementation they can lose instances
when their additional reasoning is not beneficial. The ILP configuration solves
14 instances that the baseline misses, but loses 20 instances, mostly because
SCIP or CP-SAT can become too expensive on large formulas. Similarly, the
lookahead configuration solves 11 instances that the ILP configuration misses,
but loses 7 instances because the lookahead phase can take too long. In the
current implementation, lookahead is controlled by SAT-conflict limits, but these
limits do not always bound wall-clock overhead. These losses are therefore not
fundamental limitations of the techniques, but missing guardrails in the current
implementation.

% The VBS result gives a concrete target for the next implementation iteration.
% Since ILP and lookahead build on top of the baseline, conservative
% instance-level policies should allow \coreforge{} to keep most of their gains
% while avoiding cases where their overhead dominates. For ILP, the solver should
% skip SCIP and CP-SAT, or give them a much smaller budget, on formulas that are
% too large or when early presolve does not produce enough simplification. For
% lookahead, a direct wall-clock budget should be added in addition to the
% existing per-call conflict limits. These changes should move the best individual
% configuration substantially closer to the VBS, increasing the number of solved
% instances toward 332 on this subset.

Overall, the VBS result gives a concrete target for the next implementation
iteration. Since ILP and lookahead build on top of the baseline, conservative
instance-level policies should allow \coreforge{} to keep most of their gains
while avoiding cases where their overhead dominates. For ILP, the solver should
skip SCIP and CP-SAT, or give them a much smaller budget, on formulas that are
too large or when early presolve does not produce enough simplification. For
lookahead, a direct wall-clock budget should be added in addition to the
existing per-call conflict limits. These changes should move the best individual
configuration substantially closer to the VBS. Moreover, even with these current
limitations, \coreforge{} already solves more instances than several established
MaxSAT solvers in the comparison.
\end{document}